\documentclass{PoS}
 
\title{Diffractive mechanisms in $pp \to pp \pi^{0}$ reaction at high energies}

\ShortTitle{Diffractive mechanisms in $pp \to pp \pi^{0}$ reaction at high energies}

\author{\speaker{Piotr LEBIEDOWICZ}%
\thanks{This work was supported by the Polish National Science Centre 
(on the basis of decision No.~DEC-2011/01/N/ST2/04116
and DEC-2011/01/B/ST2/04535).}\\
Institute of Nuclear Physics PAN, PL-31-342 Cracow, Poland\\
E-mail: \email{Piotr.Lebiedowicz@ifj.edu.pl}} 

\author{Antoni SZCZUREK\\
University of Rzesz\'ow, PL-35-959 Rzesz\'ow, Poland, and\\
Institute of Nuclear Physics PAN, PL-31-342 Cracow, Poland\\
E-mail: \email{Antoni.Szczurek@ifj.edu.pl}}

\abstract{
We present a study of exclusive production of $\pi^{0}$ meson in proton-proton collisions 
at high energies.
Both diffractive bremsstrahlung (Drell-Hiida-Deck type model),
photon-photon, photon-omega and photon-odderon exchange mechanisms are included in the calculation.
The $\pi^{0}$-bremsstrahlung contribution dominates at large 
(forward, backward) pion rapidities and contributes at small $\pi^0 p$ invariant mass
and could be therefore misinterpreted as the Roper resonance $N^{*}(1440)$.
Large cross sections of the order of mb are predicted.
We predict strong dependence of the slope in $t$ 
(squared four-momentum transfer between ingoing and outgoing proton)
on the mass of the supplementary excited $\pi^{0} p$ system.
At high energy and midrapidity, the photon-photon contribution 
dominates over the diffractive components, 
however, the corresponding cross section is rather small.
The photon-odderon and odderon-photon contributions are included
in addition and first estimates (upper limits) of their contributions are presented.
We suggest a search for the odderon contribution at midrapidity and
at $p_{\perp,\pi^{0}} \sim$ 0.5 GeV.
Our predictions are ready for verification at LHC.
The bremsstrahlung mechanisms discussed here contribute also to the $pp \to p(n \pi^{+})$ reaction. 
Both channels give a sizable contribution to the low-mass 
single diffractive cross section and must be included in extrapolating
the measured experimental single diffractive cross section.}

\FullConference{XXI International Workshop on Deep-Inelastic Scattering and Related Subject -DIS2013,\\
		22-26 April 2013\\
		Marseilles,France}

\begin{document}

\section{Introduction}

We discuss the exclusive $p p \to p p \pi^0$ process at high energies, see \cite{LS_pi0}.
\footnote{The exclusive pion production mechanism
is similar to $p p \to p p \omega$ \cite{CLSS} 
and $p p \to p p \gamma$ \cite{LS13} processes.}
This reaction was measured in detail only near to the pion threshold.
A nice summary of the intermediate energy data can be found in Ref.\cite{Dahl_Jensen}.
In this region of energy the corresponding cross section systematically
decreases which is consistent with the meson exchange picture.
The $p p \to p (n \pi^+)$ and $n p \to (p \pi^-) p$ reactions were measured 
at ISR and Fermilab \cite{data}.
As discussed in \cite{reports} the dominant hadronic bremsstrahlung-type 
mechanism is the Drell-Hiida-Deck mechanism
for diffractive production of $\pi N$ final states in $NN$ collisions 
\cite{Deck}.
The $p p \to p p \pi^0$ process can be measured 
with the help of the forward detectors at the LHC.
\footnote{In the past we have studied exclusive pion/kaon pairs in 
the $pp \to pp \pi^{+}\pi^{-}$ \cite{LS10, SLTCS11},
$pp \to nn \pi^{+}\pi^{+}$ \cite{LS11}
and $pp \to pp K^{+}K^{-}$ \cite{LS12} processes.
There pomeron-pomeron and pomeron-reggeon exchanges are the dominant mechanisms.
Such processes could be measured with the help of the main ATLAS or CMS
detectors (for charged pions/kaons detection), ALFA \cite{ALFA}
or TOTEM \cite{TOTEM} detectors (for protons tagging),
and the Zero Degree Calorimeters (ZDCs) \cite{ZDC} (for neutrons detection).
}

In \cite{WA102_PLB427, kirk00} a study of pseudoscalar mesons produced centrally
by the WA102 Collaboration at $\sqrt{s} = 29.1$~GeV was performed.
The results show that the $\eta$ and $\eta'$ mesons appear to have a similar
production mechanism which considerably differs from that for the 
$\pi^{0}$ production \cite{WA102_PLB427}. 
The WA102 Collaboration concentrated on central production of mesons 
and eliminates contribution comes from diffractive mechanisms discussed in \cite{LS_pi0}.
Reactions of this type $pp \to p M p$ are expected to be mediated by
double exchange processes.
For instance, the $\eta$ and $\eta'$ mesons are produced dominantly
by a mixture of pomeron-pomeron, reggeon-pomeron, and reggeon-reggeon exchanges (see \cite{LNS13}).
For the central exclusive $\pi^{0}$ production at intermediate energies 
the $\rho$-$\omega$, $\rho$-$a_{2}$ exchanges may be the dominant mechanism.

The $\pi^{0}$ can be also produced by $\gamma\gamma$, $\gamma\omega$, and $\gamma \mathcal{O}$ exchanges.
The search performed at HERA \cite{H1} 
was negative and found only an upper limit for this process 
$\sigma_{\gamma p \to \pi^0 p} < 49$~nb.
In Refs.~\cite{KN, BDDKNR} the authors discussed some results 
of exclusive pseudoscalar meson production in high energy $ep$ scattering.
As shown in Ref.\cite{BDDKNR, DDN06} the photon exchange is larger than the odderon exchange 
only at very small transverse momenta of $\pi^0$.
Here we shall consider the odderon exchange processes
using a simple phenomenological approach
and estimate their contributions in the $pp$-collisions.
\section{Sketch of formalism}
\begin{figure}
\begin{center}
(a)\includegraphics[width=.2\textwidth]{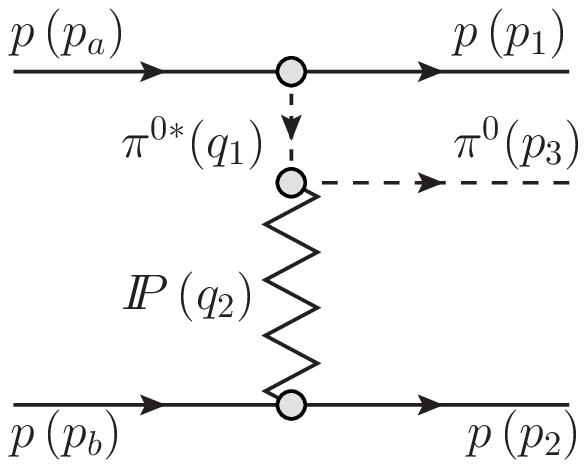} \quad
(b)\includegraphics[width=.2\textwidth]{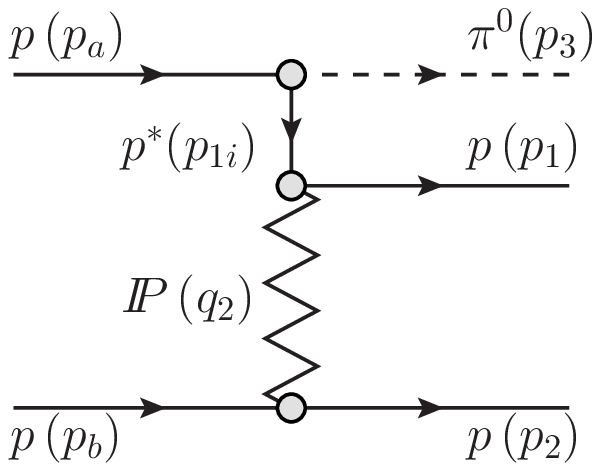} \quad
(c)\includegraphics[width=.2\textwidth]{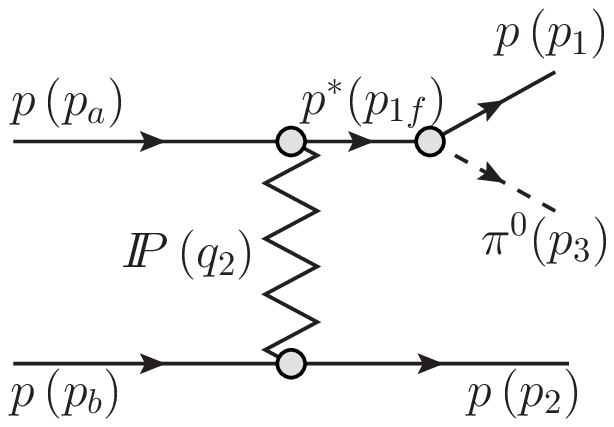} \quad
\end{center}
\caption{
Diagrams of the $\pi^0$-bremsstrahlung amplitudes driven by
the pomeron ($I\!\!P$) exchange in the $pp$-collisions:
(a) pion exchange, (b) proton exchange, and (c) direct production.
}
\label{fig:diagrams_deck}
\end{figure}
The dominant diffractive bremsstrahlung mechanisms for the exclusive $p p \to p p \pi^0$ reaction 
driven by pomeron exchange are depicted in Fig.\ref{fig:diagrams_deck}.
The formalism and more details has been shown and discussed elsewhere \cite{LS_pi0}.
In our non-resonant model we take into account also absorption effects, see Section II~B of \cite{LS_pi0}.
\footnote{
There are also resonance contributions, due to 
diffractive excitation of some nucleon resonances
and their subsequent decays into the $p + \pi^0$ ($\bar p + \pi^0$) channels.
The dominant contributions are due to $N^{*}$ resonant states being
members of the nucleon trajectory (the $N^{*}(1680)$~5/2$^+$ state). 
Although a huge contribution of the Roper resonance $N^{*}(1440)$
was suggested recently \cite{JKOS2012}, as discussed in \cite{LS_pi0},
their contribution may be to some extent an artifact of a fit 
which does not include the $\pi^{0}$-bremmstrahlung mechanism.}
\begin{figure}   
\begin{center}
(a)\includegraphics[width=.2\textwidth]{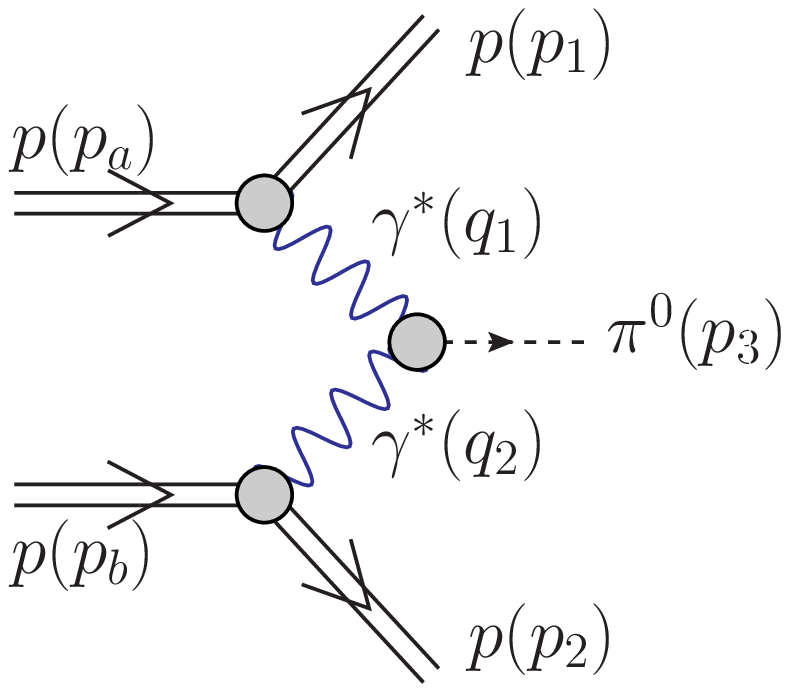} \quad
(b)\includegraphics[width=.2\textwidth]{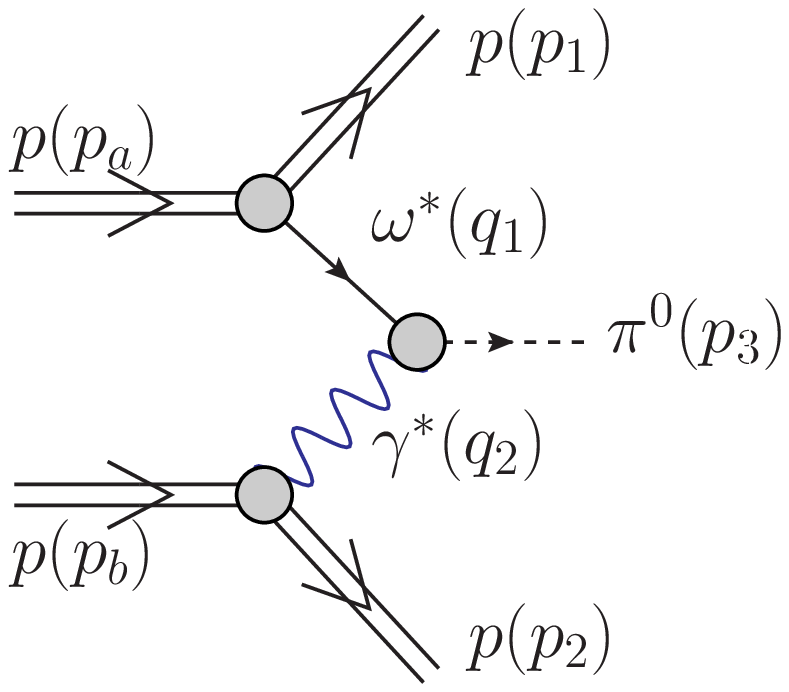} \quad
(c)\includegraphics[width=.2\textwidth]{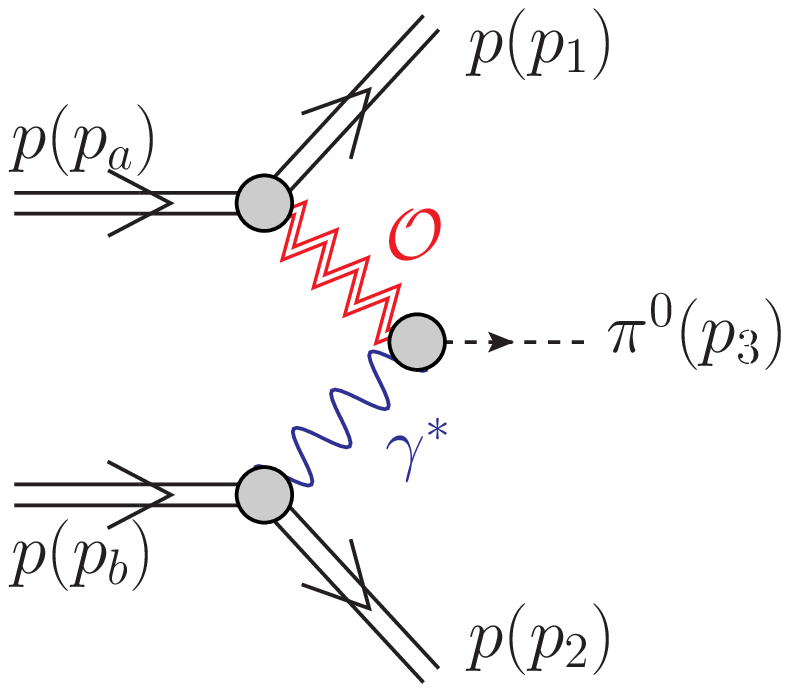} 
\end{center}
\caption{
A sketch of the photon-photon (a), photon-omega meson (b), and photon-odderon (c) exchanges
in the $pp \to pp\pi^0$ reaction.}
\label{fig:diagram_gamgam_pi0}
\end{figure}
The new mechanisms, never discussed so far in the literature,
are shown schematically in Fig.~\ref{fig:diagram_gamgam_pi0}.

\section{Results}
\label{section:Results}
\begin{figure} 
\centering
\includegraphics[width=0.64\textwidth]{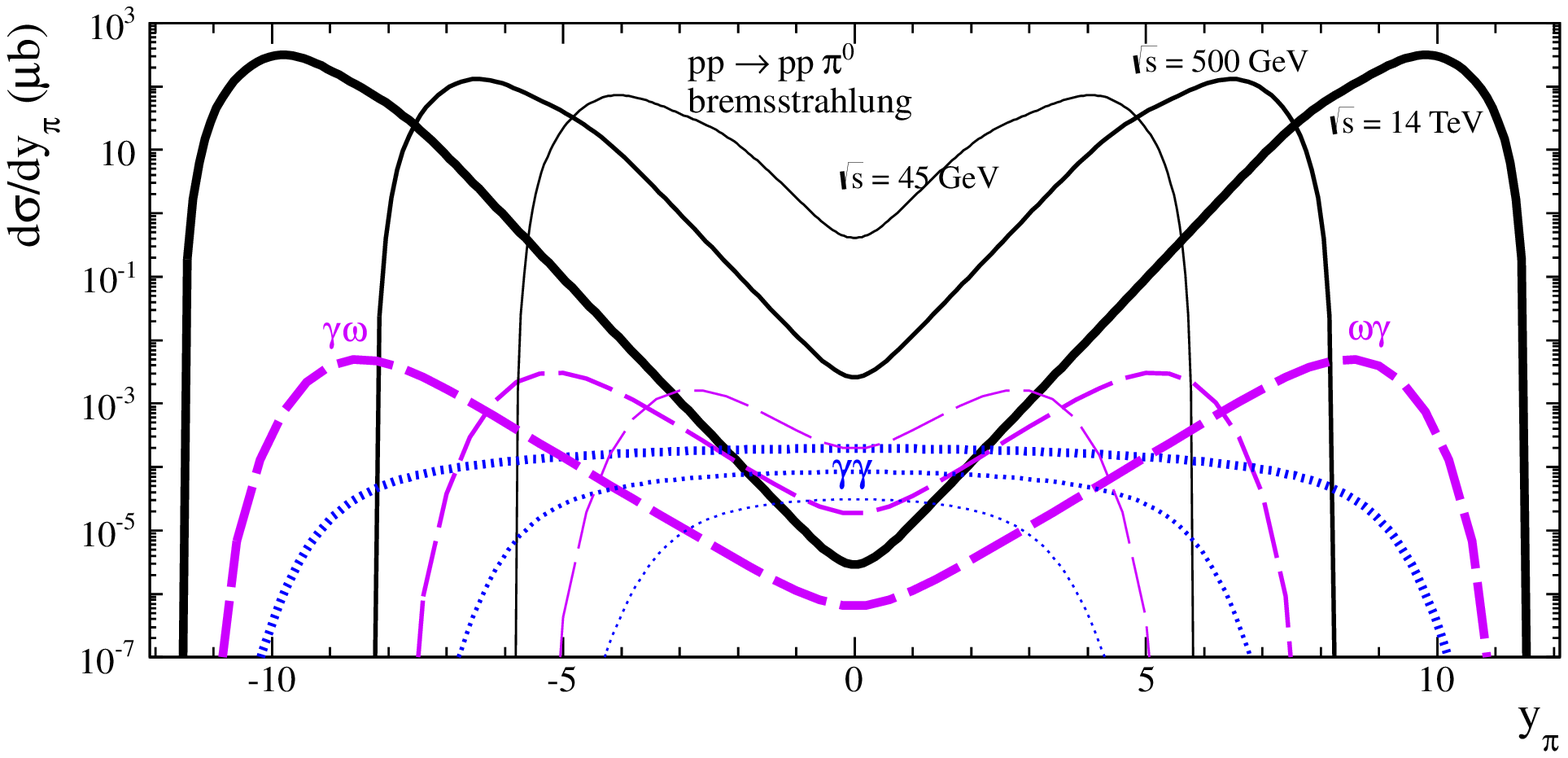}
\includegraphics[width=0.325\textwidth]{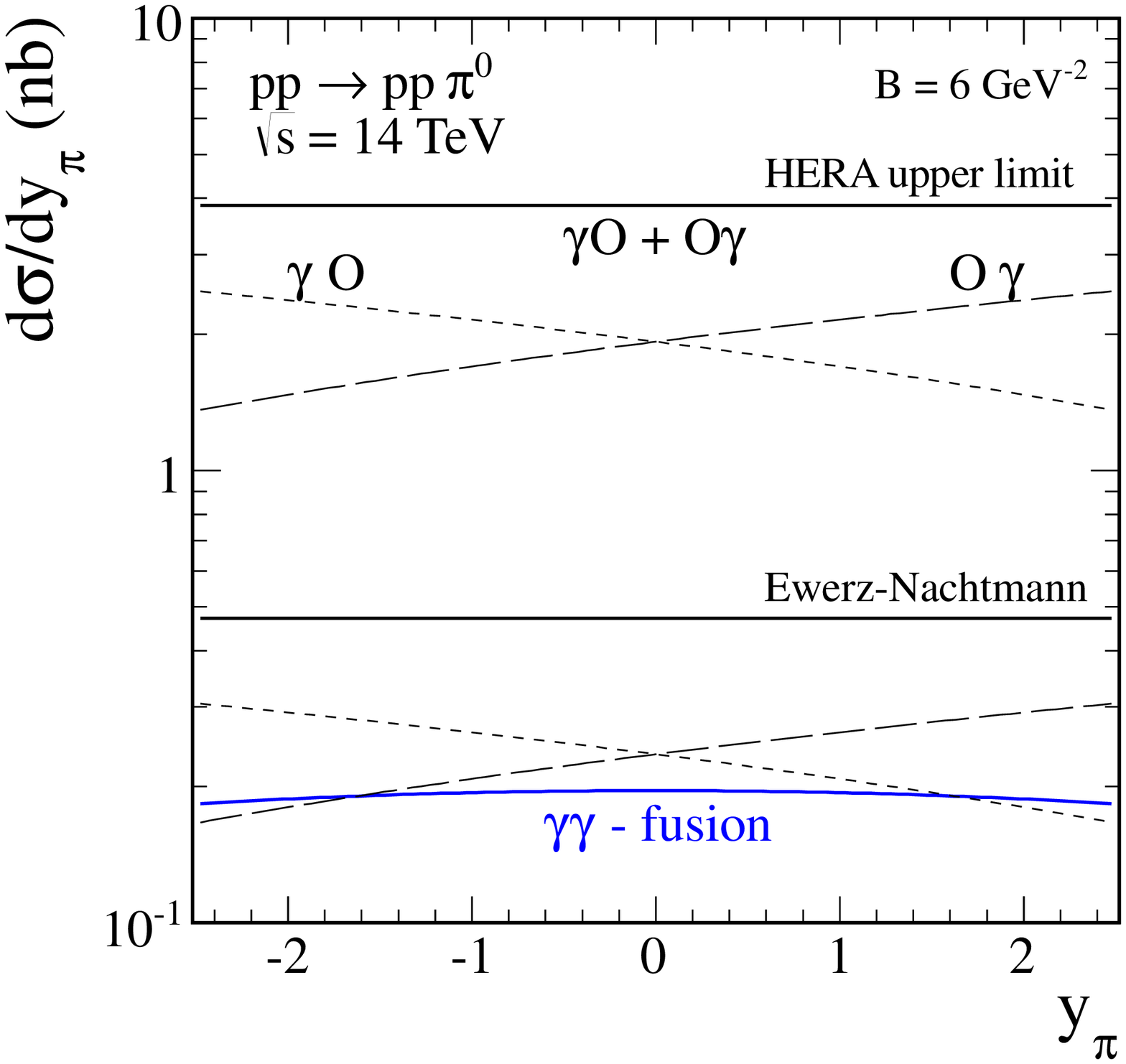}
\caption{
The Born distributions in rapidity of $\pi^0$ 
at $\sqrt{s}=45$~GeV (ISR), 500~GeV (RHIC), and 14~TeV (LHC).
The $\pi^{0}$-bremsstrahlung contribution (black solid lines) and
$\omega \gamma$ ($\gamma \omega$) exchanges (violet dashed lines) 
peaks at forward (backward) region of rapidity, respectively,
while $\gamma \gamma$ fusion (blue dotted lines) contributes at midrapidity.
In this calculation we have used 
$\Lambda_{N} = \Lambda_{\pi} = 1$~GeV of the hadronic form factors.
In the right panel, in addition, we show 
results from $\gamma \mathcal{O}$- and $\mathcal{O} \gamma$-fusion
for two estimates of the $\gamma p \to \pi^0 p$ cross section (energy independent).
Individual contributions of photon-odderon (short dashed line)
and odderon-photon (long dashed line) are shown separately.}
\label{fig:dsig_dy_all}
\end{figure}

Now we present some distributions
of the exclusive $\pi^{0}$ meson production in proton-proton collisions
for all considered processes from Figs.\ref{fig:diagrams_deck} and \ref{fig:diagram_gamgam_pi0}.
The rapidity distributions of $\pi^{0}$
are shown in Fig.\ref{fig:dsig_dy_all} (left panel)
at $\sqrt{s}=45$~GeV (ISR), 500~GeV (RHIC) and 14~TeV (LHC).
If energy increases the dominant $\pi^{0}$-bremsstrahlung contributions
(solid lines) and omega-photon (photon-omega) exchange processes
(dashed lines) become better separated and occupy forward/backward region of rapidity.
The photon-photon fusion contribution (dotted lines)
dominates over the diffractive mechanism
at the LHC energy and midrapidity region $-2 < y_{\pi^{0}} < 2$.
In the right panel we show predictions of the odderon contributions
for the HERA upper limit ($\sigma_{\gamma p \to \pi^{0} p} = 49$~nb)
and for the Ewerz-Nachtmann estimate ($\sigma_{\gamma p \to \pi^{0} p} = 6$~nb)
at $\sqrt{s} = 14$~TeV.
The total cross section in the first case
is less than 20~nb in the rapidity region $|y_{\pi^{0}}| < 2.5$.
The corresponding curve is more than an order
of magnitude larger than the photon-photon contribution.
The cut on meson $p_{\perp,\pi^{0}}$ should enhance the relative odderon contribution
(see \cite{LS_pi0} for more details).

\begin{figure} 
\centering
\includegraphics[width=0.325\textwidth]{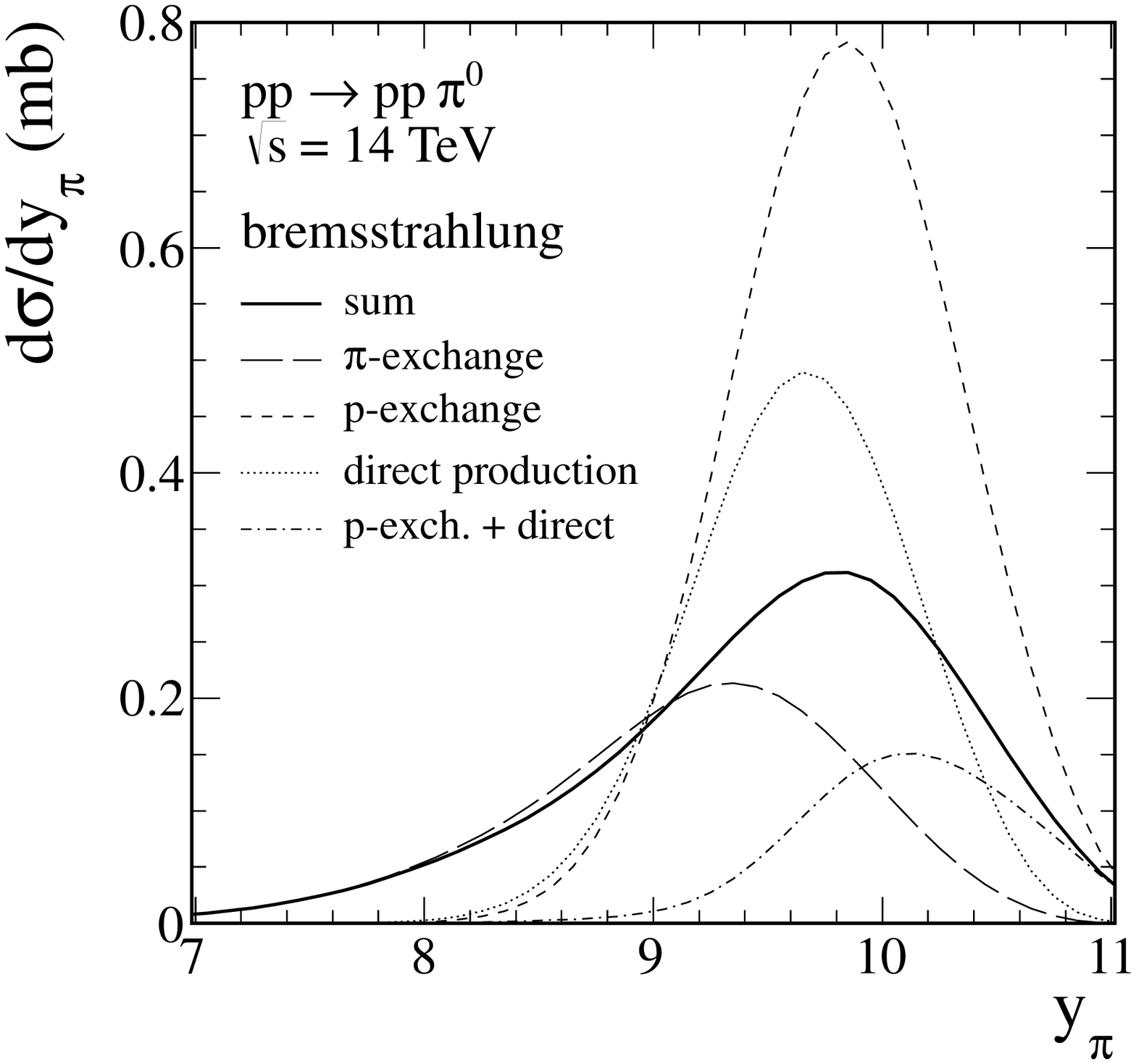}
\includegraphics[width=0.325\textwidth]{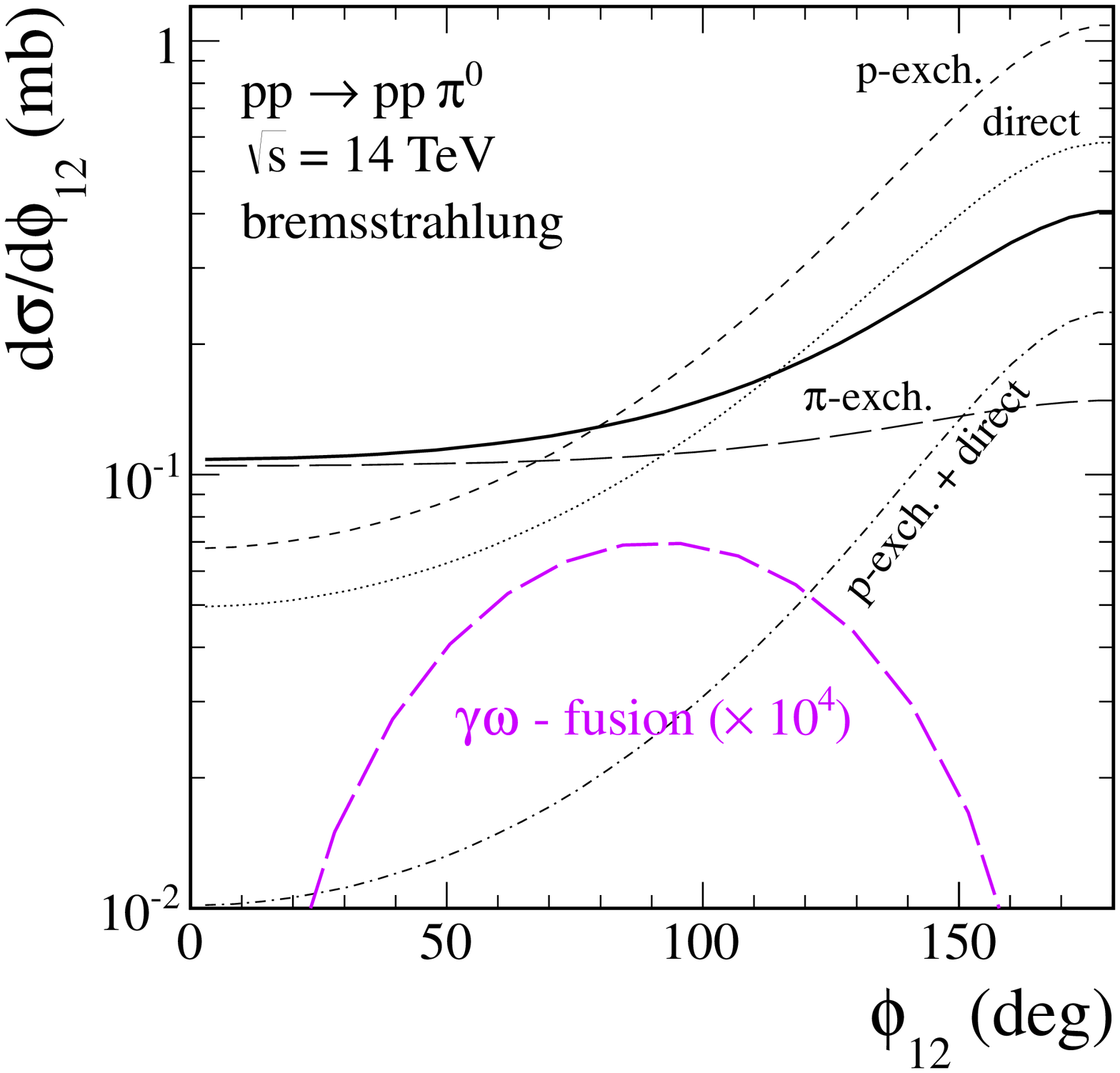}
\includegraphics[width=0.325\textwidth]{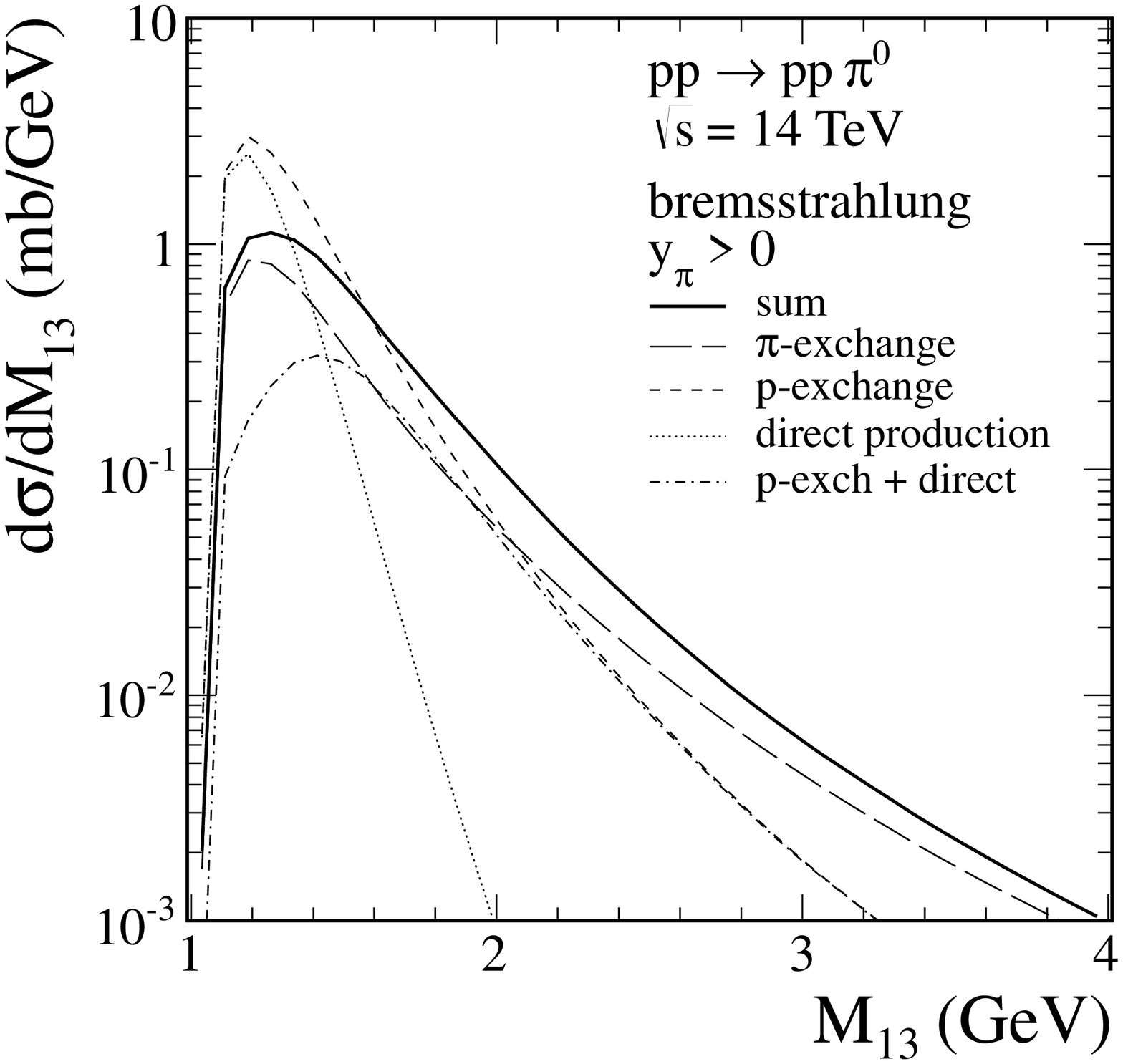}\\
\includegraphics[width=0.325\textwidth]{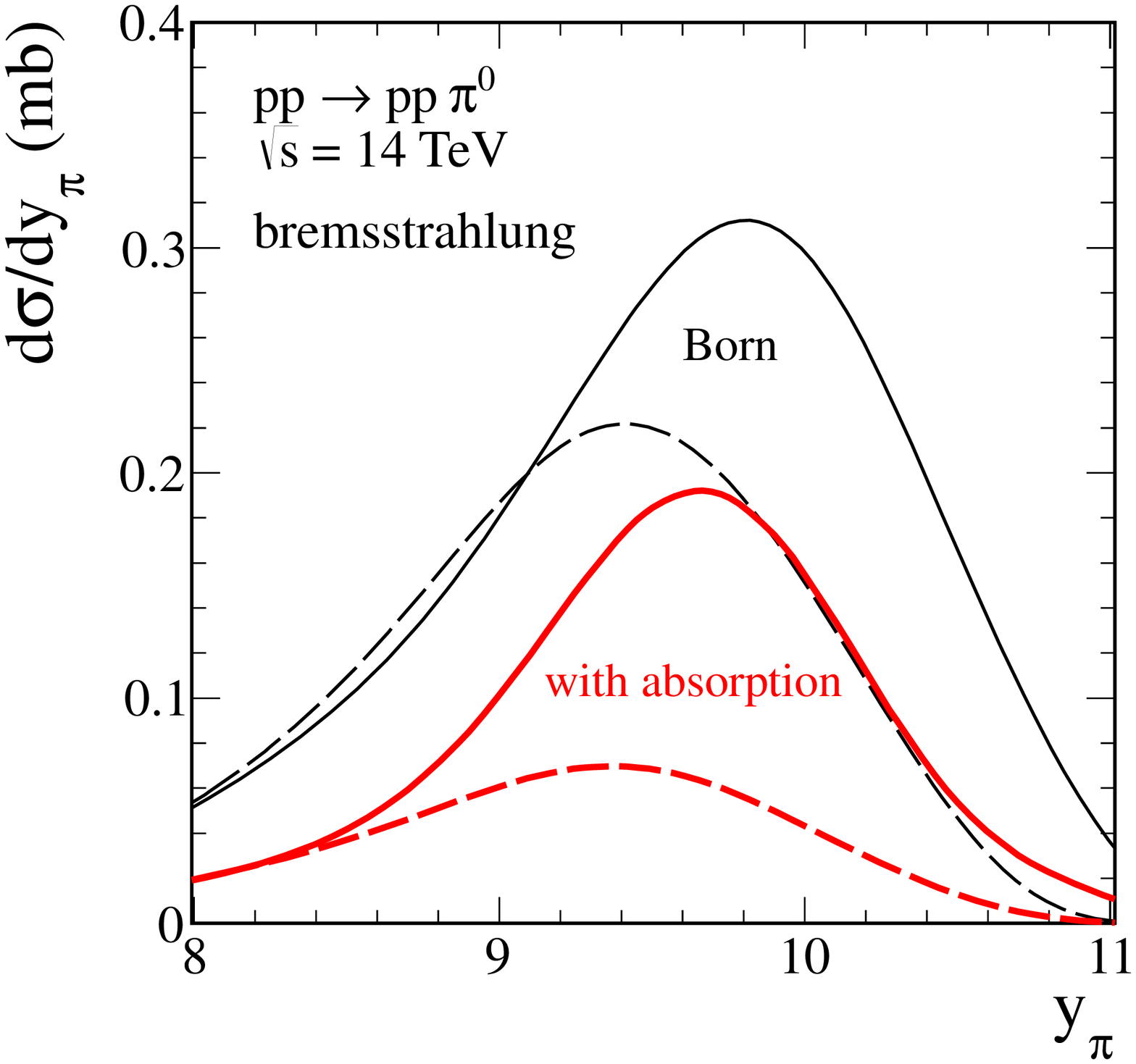}
\includegraphics[width=0.325\textwidth]{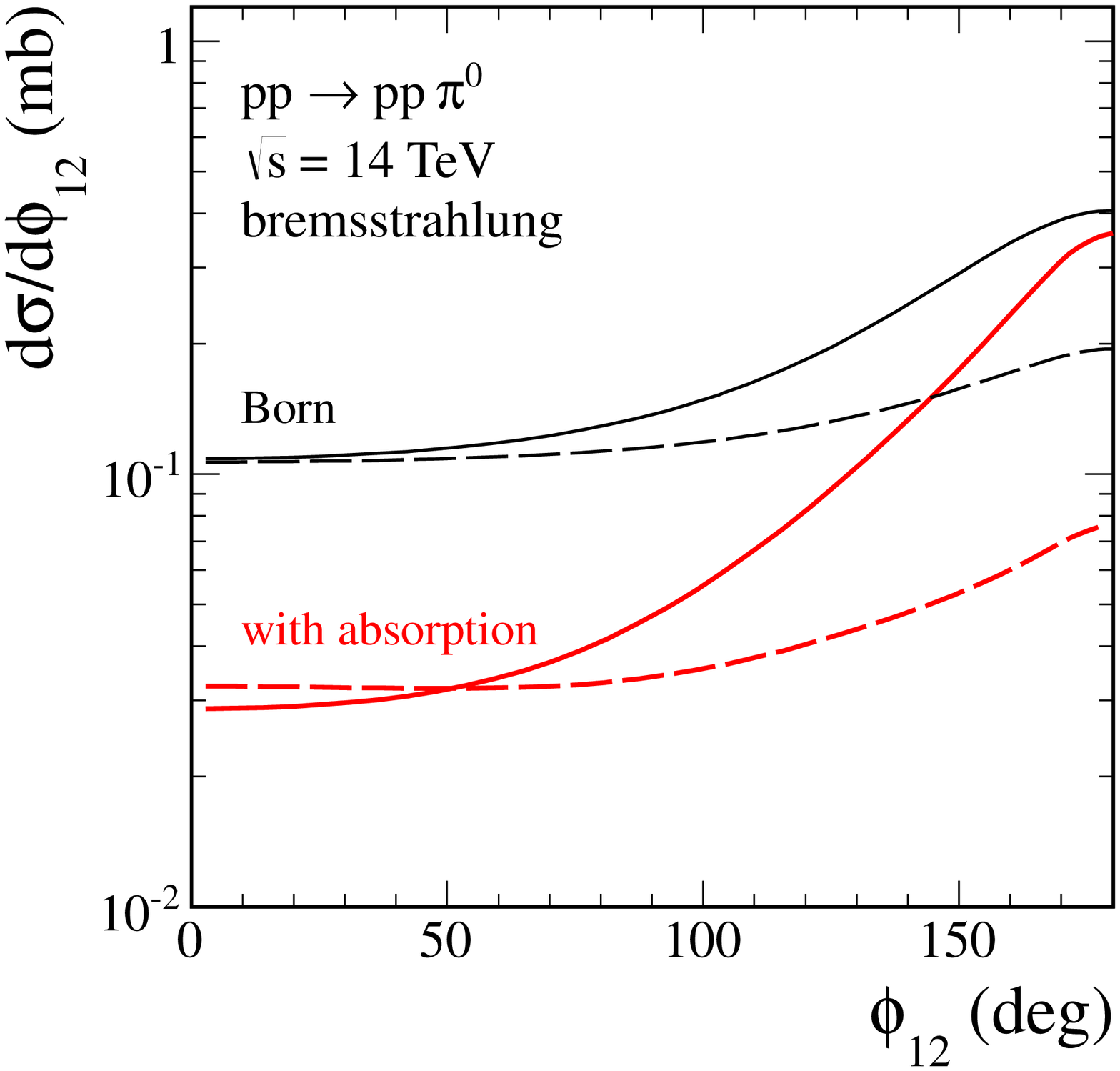}
\includegraphics[width=0.325\textwidth]{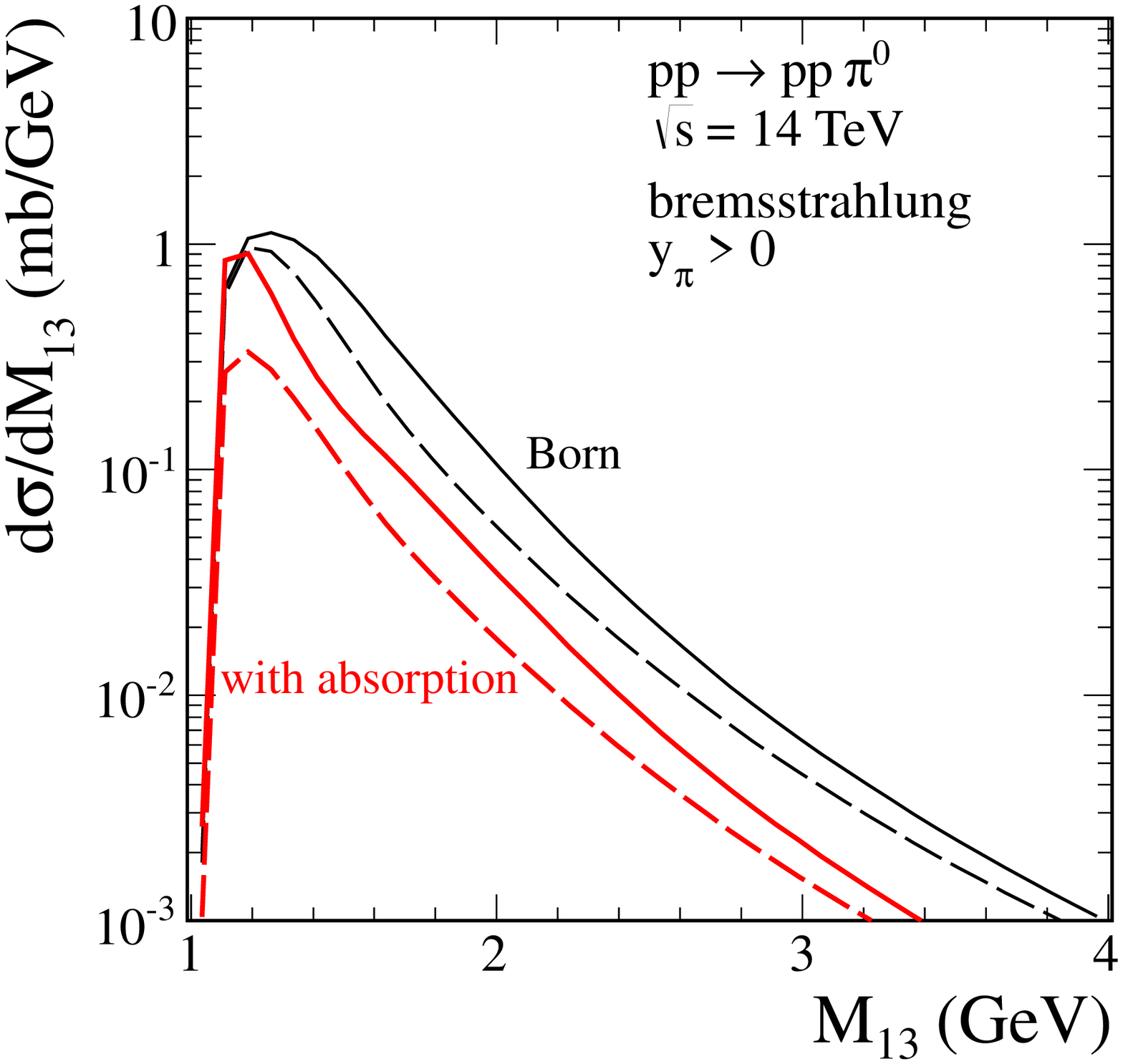}\\
\caption{Distributions in $\pi^{0}$ rapidity, relative azimuthal angle between outgoing protons
and $p \pi^{0}$ invariant mass for the $\pi^{0}$-bremsstrahlung mechanism at $\sqrt{s} = 14$~TeV.
The individual contributions to the cross section are shown separately.
Results of calculations without absorption effects (upper black lines)
and with absorption effects (lower red lines) are presented.
The solid lines correspond to the form factor parameters $\Lambda_{N} = \Lambda_{\pi} = 1$~GeV
while the dashed lines are for $\Lambda_{N} = 0.6$~GeV and $\Lambda_{\pi} = 1$~GeV.}
\label{fig:dsig_dy_deco}
\end{figure}

In Fig.\ref{fig:dsig_dy_deco} we present differential distributions
in $\pi^{0}$ rapidity, relative azimuthal angle between outgoing protons
and $p \pi^{0}$ invariant mass for the $\pi^{0}$-bremsstrahlung mechanism.
The difference between the solid ($\Lambda_{N} = \Lambda_{\pi} = 1$~GeV)
and long-dashed ($\Lambda_{N} = 0.6$~GeV and $\Lambda_{\pi} = 1$~GeV)
curves represents the uncertainties on the hadronic form factors; see \cite{LS_pi0}.
In addition, we show the individual contributions of diagrams given in 
Fig.\ref{fig:diagrams_deck} to the cross section.
We observe a large cancellation between the two terms in the amplitude
(between the initial ($p$-exchange) and the final state radiation (direct production)).
Absorptive corrections cause damping of the cross section
by a factor 2 to 3 and modify the shape of some distributions.
As can be seen in the middle panel the diffractive contribution is peaked 
at the back-to-back configuration ($\phi_{12} = \pi$)
in contrast to the $\gamma \gamma$- or $\gamma \omega$-fusion.
The discussed $pp \to pp \pi^{0}$ process gives a sizeable
contribution to the low mass single diffractive cross section.

\begin{figure}
\centering
\includegraphics[width = 4.8cm]{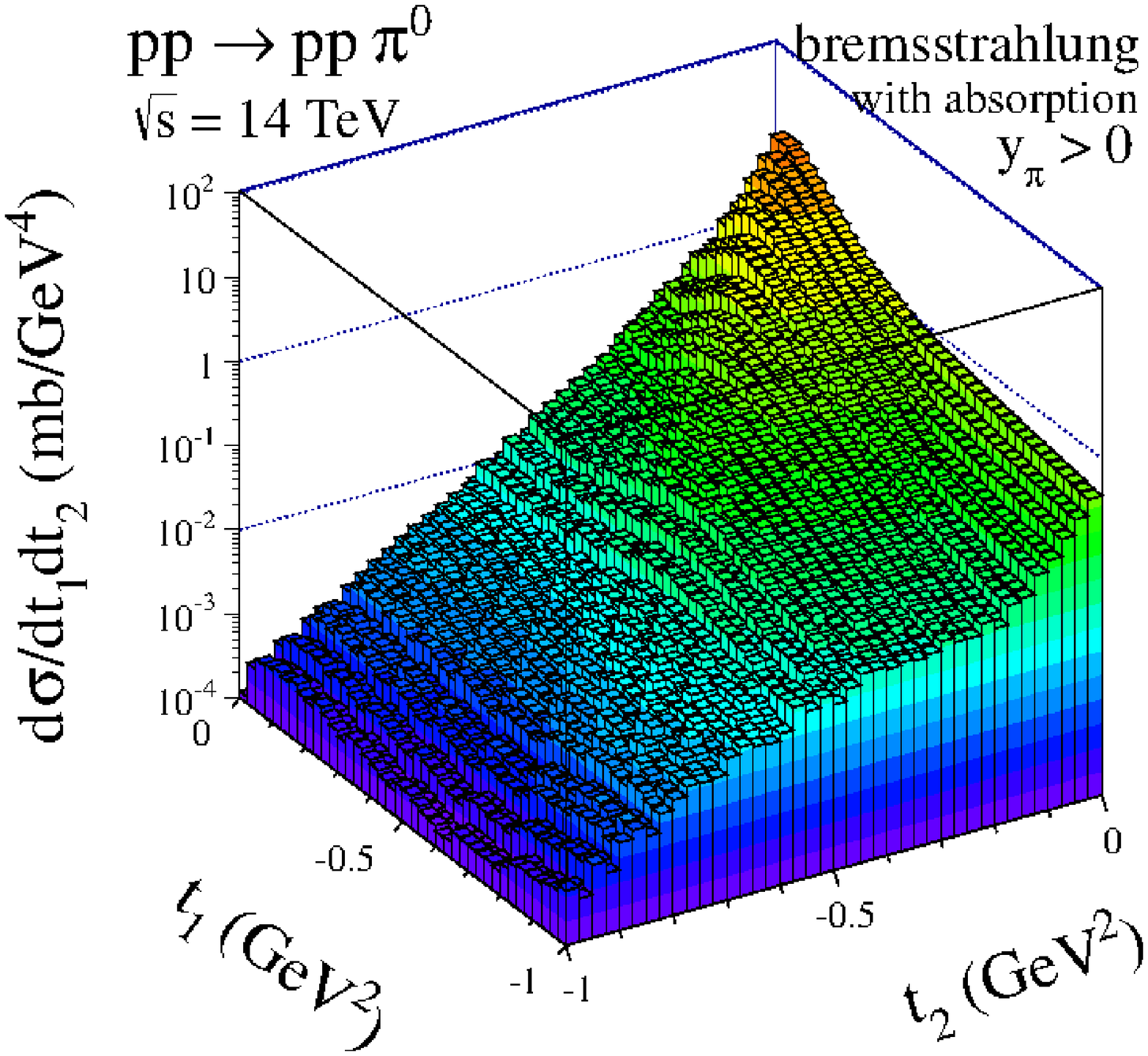}
\includegraphics[width = 4.8cm]{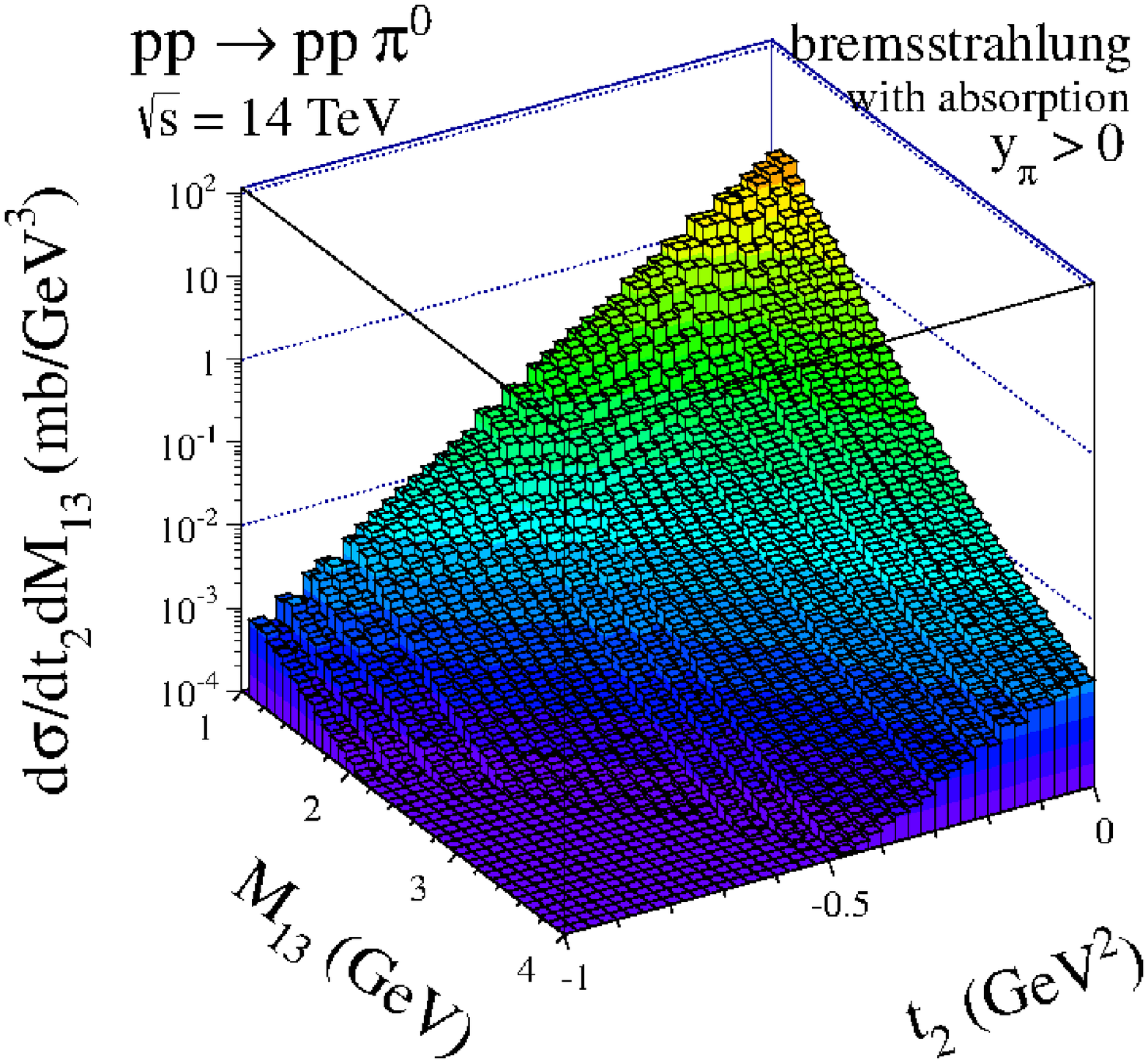}
\includegraphics[width = 4.8cm]{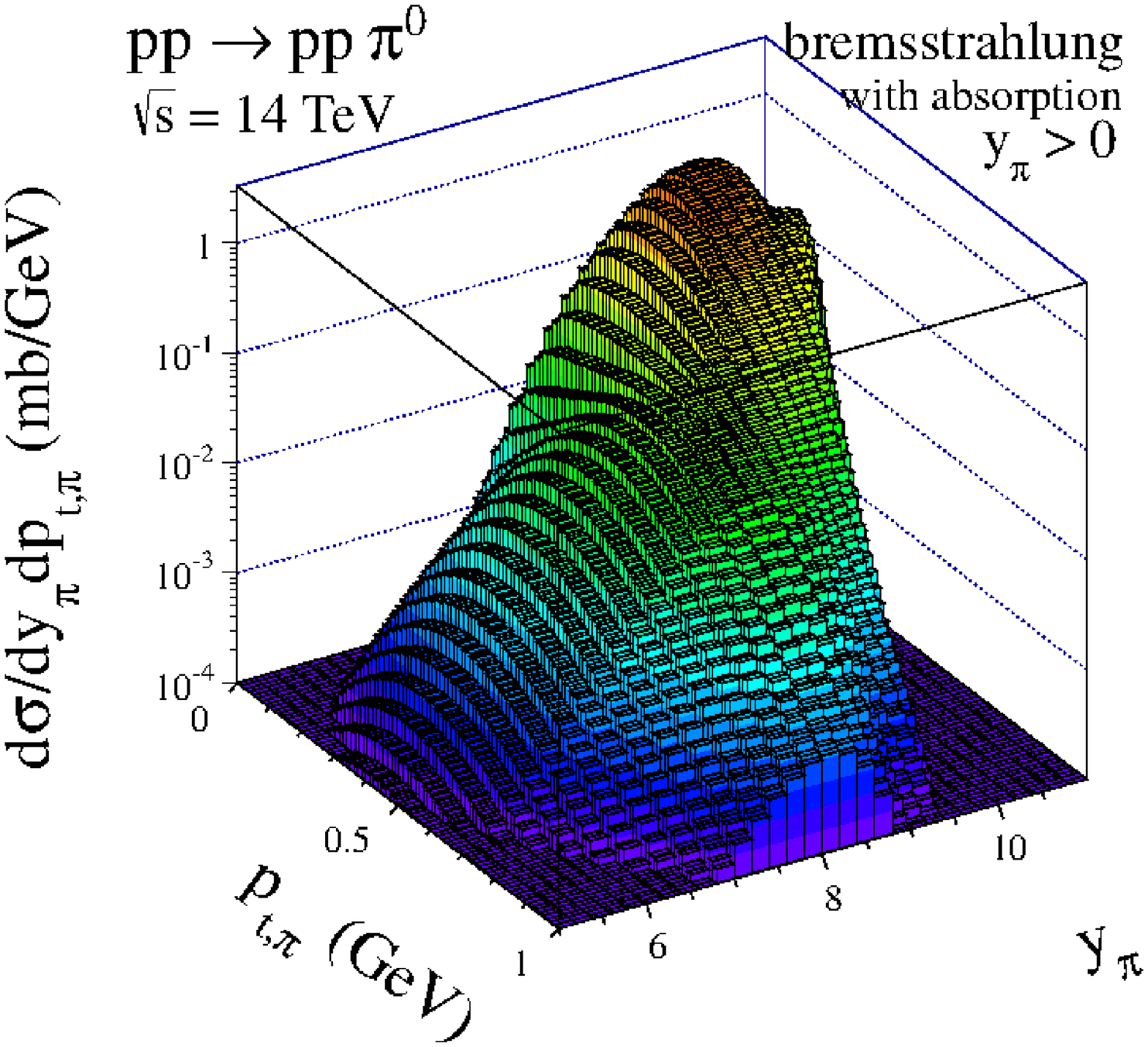}
\caption{
Distributions in ($t_{1}, t_{2}$), ($t_{2}, M_{13}$) and ($y_{\pi^{0}}, p_{\perp,\pi^{0}}$) spaces
at $\sqrt{s} = 14$~TeV for the $\pi^{0}$-bremsstrahlung contribution
in forward region ($y_{\pi^{0}}>0$).
Here absorptive corrections are included and $\Lambda_{N} = \Lambda_{\pi} = 1$~GeV.}
\label{fig:map_t1t2}
\end{figure}

In Fig.\ref{fig:map_t1t2} we show some two-dimensional distributions
for the diffractive contribution at $\sqrt{s} = 14$~TeV with absorptive corrections.
The distributions in $t_{1}$ or $t_{2}$ (left panel) 
are different because we have limited to the case of $y_{\pi^{0}} > 0$ only.
The $t$-distributions discussed here could in principle be obtained with 
the TOTEM detector at CMS to supplement the ZDC detector for the measurement of neutral pions. 
Similar analysis could be done by the ALFA detector for proton tagging at ATLAS.
One can observe different behavior of slope in $t_{2}$ 
for different masses of the $p \pi^{0}$ system (middle panel).
A similar effect was observed for $pp \to p(n \pi^{+})$
and $np \to (p \pi^{-})p$ reactions at much lower energies \cite{data}.
In right panel the sizeable correlations between pion rapidity and transverse momentum
can be observed which is partially due to interference of different amplitudes.

\section{Conclusions}
We have included the $\pi^{0}$-bremsstrahlung 
from the initial or final proton, diffractive $\pi^0$-rescattering,
photon-photon, photon-omega and photon-odderon fusion processes.
We have found very large cross sections of the order of mb.
The dominant contributions are placed at large rapidities
where they could be measured with the help of the forward detectors at the LHC.
The total (integrated over phase space) cross section is almost energy independent.
Absorptive effects lower the cross section by a factor 2 to 3; see Table~I in \cite{LS_pi0}.
At the LHC the two-photon fusion mechanism ``wins" with
the diffractive mechanisms at midrapidity.
However, the transverse momenta of neutral pions in this region
are very small and therefore such pions are very difficult to measure.
The $\gamma \omega$ or $\omega \gamma$ exchanges have been found to be 
significant only in backward or forward rapidities, respectively, 
and are small at midrapidities due to $\omega$ reggeization. 

We have also presented first estimates of the photon-odderon and
odderon-photon contributions based on the upper limit of
the $\gamma p \to \pi^0 p$ cross section obtained at the HERA
as well as estimates based on a nonperturbative approach of Ewerz 
and Nachtmann (\cite{DDN06}, \cite{EN2007}) which makes use of chiral symmetry and PCAC. 
Based on the HERA upper limit we conclude that the cross section for 
the contribution to the $p p \to p p \pi^0$ reaction
is smaller than 20~nb in the rapidity region $|y_{\pi^{0}}| < 2.5$.
Any deviation from the $\gamma \gamma \to \pi^0$ contribution
to transverse momentum distribution of neutral pions at midrapidity 
would be a potential signal of photon-odderon (odderon-photon) contributions. 
One can expect potential deviations from the photon-photon
contribution at $p_{\perp,\pi^{0}} \sim 0.5$~GeV. 
This requires dedicated studies if the considered process could be measured at the LHC.

We have calculated differential cross sections
for the exclusive $p p \to p p \pi^0$ reaction at high energies.
If one limits to separate regions of $y_{\pi^0} < 0$ or $y_{\pi^0} > 0$
(one-side excitation), then the distributions in proton transverse
momenta $t_{1}$ and $t_{2}$ are quite different -- one reflecting 
the pion/nucleon exchange and the second reflecting the pomeron exchange. 
The same is true for the $p_{\perp,1}$ and $p_{\perp,2}$ distributions, see \cite{LS_pi0}. 
Analysis of such details would be a useful test of the model. 
The distribution in the mass of the excited $\pi^0 p$ system 
peaks at small $M_{\pi p}$ and quickly drops when the mass increases.
Such a distribution reminds the spectral shape of the Roper resonance 
fitted recently to an old single-diffractive data.
We have obtained an interesting correlation between
the mass of the excited system and the slope of the $t$ distributions
well represented in a two-dimensional plot $d \sigma /dt dM (t,M)$.
Particularly interesting is the distribution in azimuthal angle
between outgoing protons.
The distribution has a maximum at relative angle $\phi_{12} = \pi$.
The sensitive nature of the cancellation between 
proton-exchange and direct production amplitudes 
leads to a situation where minor changes in the parametrizations of these amplitudes 
can have large effects on discussed distributions.


The reaction discussed here is interesting also in a much broader context.
First of all, it may constitute a sizable fraction of 
the pion inclusive cross section at very forward/backward (pseudo)rapidities. 
Second, it leads to a production of very energetic photons 
($\sim 0.5-2$~TeV) from the decay of the forward $\pi^0$'s.
Third, the $\pi^{0}$-bremmstrahlung mechanism sizably contributes to the single diffractive
cross section and as a consequence to the total inelastic cross section.
This contribution is not included in any existing Monte Carlo code.



\end{document}